# Artificial Intelligence Based Cloud Distributor (AI-CD):

# Probing Low Cloud Distribution with a Conditional Generative Adversarial Network

Tianle Yuan[1,2,3]

[1]Earth Sciences Directorate, NASA Goddard Space Flight Center
[2]Joint Center for Earth Systems Technology, University of Maryland, Baltimore County
[3]Benyuan Tech, Rockville, Maryland

Key points:
- We train a generative adversarial network (cGAN) model to generate spatial distribution of low clouds conditioned on meteorological variables.
- We call the approach Artificial Intelligence-based Cloud Distributor (AI-CD) and it can generate diverse and realistic cloud scenes given large-scale meteorological variables.
- The AI-CD approach captures the relationship between large-scale variables such as estimated inversion strength, sea surface temperature, and subsidence rate and cloud properties.

Abstract:

Here we introduce the artificial intelligence-based cloud distributor (AI-CD) approach to generate two-dimensional (2D) marine low cloud reflectance fields. AI-CD uses a conditional generative adversarial net (cGAN) framework to model distribution of 2-D cloud reflectance in nature as observed by the MODerate resolution Imaging Spectrometer (MODIS). Specifically, the AI-CD models the conditional distribution of cloud reflectance fields given a set of large-scale environmental conditions such as instantaneous sea surface temperature, estimated inversion strength, surface wind speed, relative humidity and large-scale subsidence rate together with random noise. We show that AI-CD can not only generate realistic cloudy scenes but also capture known, physical dependence of cloud properties on large-scale variables. AI-CD is stochastic in nature because generated cloud fields are influenced by random noise. Therefore, given a fixed set of large-scale variables, an ensemble of cloud reflectance fields can be generated using AI-CD. We suggest that AI-CD approach can be used as a data driven framework for stochastic cloud parameterization because it can realistically model sub-grid cloud distributions and their sensitivity to meteorological variables.

Text
1. Introduction

Weather and climate models have finite resolution and processes that are characterized by scales smaller than the grid size have to be parameterized using resolved scale variables through schemes inspired by our physical understanding and statistical regression(Arakawa, 2004). Such schemes have enjoyed success in simulating various fields, such as the global patterns of precipitation, cloud amount, and energy balance, by tuning their parameters so that models can match the climatology of their observational counterparts (Flato et al., 2013). However, such approaches have their set of limitations: the need for tuning, large errors in instantaneous values and climatology, large compensating errors, and uncertainty in their sensitivity to future climate changes.

For the particular case of low cloud modeling, systematic errors have persisted for a long time in global models. Errors in low cloud modeling have strong impacts on other critical variables such as sea surface temperature and the energy balance of the Earth system ( Wang et al., 2014). Sensitivity of low clouds to future climate change is closely related to model climate sensitivities (Bony & Dufresne, 2005). In fact, low cloud modeling is a key source of uncertainty in our ability to project future climate changes (Dufresne & Bony, 2008). The key challenge is to determine cloud properties in a grid that are results of a complex set of interactive processes operating at a range of scales (Robert Wood, 2012), e.g. from $O(\mu m)$ to $O(10^5 m)$ and therefore it is prohibitively expensive to numerically resolve these processes. A non-traditional approach to address this challenge is to run cloud resolving models or large-eddy simulations for each grid to derive statistics of low clouds (Randall et al., 2003). Recent developments of this approach include training neural networks or random forests models to approximate statistics from fine-scale models to save computing resources (O'Gorman & Dwyer, 2018; Rasp et al., 2018).

Here we propose the artificial intelligence-based cloud distributor (AI-CD) approach as new framework to address sub-grid low cloud modeling in a data-driven fashion. The AI-CD approach is also stochastic in nature. Its parameters are trained by large amount of high-resolution low cloud observations (or explicit simulations) together with coarse large-scale data. In this prototype study, we use the AI-CD to model low cloud albedo spatial distribution by using cloud reflectance as a proxy. In the following, we present data and method in section 2. Results will be shown in section 3 and discussions in section 4. We conclude in section 5.

2. Data and Method
2.1 Data
The MODerate resolution Imaging Spectroradiometer (MODIS) onboard both the Terra and the Aqua satellites has 36 spectral channels from deep blue to thermal infrared. Its measurements have been used for a wide range of applications and one of them is to observe and retrieve cloud properties(Platnick et al., 2016). We use single channel reflectance data at 0.55 µm from Aqua as our training data. We use the reflectance values as a proxy for cloud optical depth or albedo because they are directly observable and we want to avoid missing data or discontinuity created by the process of retrieving cloud optical depth or albedo. We use the 1km resolution data and apply a few simple filters to get only marine low cloud scenes. First, we remove pixels that have viewing zenith angles greater than 45 degrees to avoid edge pixels. We then use cloud phase and cloud height retrievals from the MOD06 products to keep scenes that are at least 5% covered by low clouds and whose high clouds fraction is less than 10%. Each scene is a 128x128 image and low clouds are defined as cloudy pixels whose cloud top does not exceed 4000m. We use Aqua data from January, April, July, and October of 2010 and 100,000 training images are randomly selected from a pool of 600,000 filtered scenes that come from the Northeast and Southeast Pacific, the Tropical Atlantic, and Southeast Atlantic. The data include both stratocumulus and trade cumulus regimes. We normalize the MODIS reflectance data so they have unit variance and zero mean.

For each image, we collocate eight meteorological variables with each scene. They include the estimated inversion strength (EIS), sea surface temperature (SST), surface pressure, relative humidity at 1000mb, surface winds, wind shear between surface and 700mb, and large-scale subsidence rate at 500mb. We test data from both the European Center for Medium-range Weather Forecasts Reanalysis Interim (ERA-Interim) project and the Modern-Era Retrospective analysis for Research and Applications, Version 2 (MERRA-2) of NASA. Results are similar. For each low cloud scene, we have a low cloud reflectance image and collocated eight meteorological variables.

2.2 Method

We train a conditional generative adversarial network model on the collocated dataset. Generative adversarial networks (GANs) are a class of deep learning techniques where two neural networks play a zero-sum game (Goodfellow et al., 2014). One neural network, a generator (G), generates samples from a noise vector z and the other neural network, a discriminator (D), tries to tell whether a sample is from real data or generated by G. The generator tries to generate increasingly more realistic samples to fool the discriminator while the discriminator improves itself not to be fooled. At equilibrium, G can generate samples that are inseparable from real data while the probability of D correctly separating samples from real data is 0.5, i.e. as skillful as a random guess since G can generate perfectly realistic data. The training process for a regular GAN is achieved through playing a minimax game with a value function V(D,G):

$\min_G \max_D V(D, G) = \mathbb{E}_{x \sim p_{data}(x)}[logD(x)] + \mathbb{E}_{z \sim p_z(z)}[\log(1 - D(G(z)))].$

In a conditional GAN (cGAN) setup (Mirza & Osindero, 2014), G generates samples based on a condition vector, y, and D is also given y as auxiliary information when attempting to separate generated samples from real data (see Figure 1 for an illustration). The objective function of the minimax game becomes:

$\min_G \max_D V(D, G) = \mathbb{E}_{x \sim p_{data}(x)}[logD(x|y)] + \mathbb{E}_{z \sim p_z(z)}[\log(1 - D(G(z|y)))].$

It is worth noting that at its theoretical limit, a well-trained G can recover the true distribution of x and therefore generate realistic samples given conditions, y. Since we train the model with observation data and condition vectors from reanalysis data, a well-trained G learns the true distribution of low clouds. During training, we alternate between training the generator and the discriminator. For our model setup, we find training the generator three times before training the discriminator once gives us the best result. We use the Adam optimizer (Kingma & Ba, 2014) and a constant batch size of 128. We use the Jensen-Shannon divergence as our GAN loss.

For the application in this paper, we aim to generate realistic samples of 2-D cloud distributions, i.e. reflectance images, given meteorological variables. We use a deep neural network for both G and D. G takes a noise vector of size 128 and eight meteorological variables to generate samples of size 128x128. It has two densely connected layers and four upsampling layers and four convolutional layers. We use rectified linear units as activation functions and add a batch normalization layer after each convolutional layer. D is also a deep convolutional neural network. It takes an input reflectance image, real or generated, together with the meteorological variables and has three convolutional layers and two dense layers. It outputs a vector indicating the probability of the input image being real or generated. Once trained, the generator will act as the AI-CD that distributes clouds spatially based on a combination of a noise vector and meteorological variables.

3. Results

Figure 2 shows a set of random samples from real MODIS data. It illustrates that low clouds can be organized into a rich set of morphologies, or spatial structures (R Wood & Hartmann, 2006). A few well-known low cloud morphologies are present in Figure 2. G7, H6, and J7 are closed-cellular stratocumulus scenes. J4 and J5 are open-cellular scenes. The most prevalent type is those that have low cloud fraction and are made up of scattered convective clouds. These often occur in the trade wind regions where shallow convections randomly pop up that have various depth and spatial forms, sometimes forming clusters. The wide range of low cloud forms and their spatial scales testify complex processes at work to create these cloud structures, which include, but not limited to, radiation, precipitation and microphysics, large-scale meteorology. They are good examples of why low clouds are hard to model and remain an uncertainty source in current climate models.

We use a cGAN to find the conditional probability distribution of cloud spatial structures in the reflectance space given a set of large-scale meteorological variables. A well-trained model is clearly of stochastic nature since eight variables are not enough to deterministically predict cloud distribution in an128x128 domain. The stochasticity should be intrinsic to a cloud parameterization scheme and our aim is that our cGAN based AI-CD, i.e. the trained generator, can capture key aspects of the physics between large-scale variables and cloud properties while still being stochastic. Figures 3-4 shows two sets of examples of the output from a trained generator. Panels in each figure are generated with a fixed noise vector that is different for different figures and they differ only in meteorological variables. There are eight rows in each figure and each represents one meteorological variable, which increases its value from left to right.

We note that for a fixed noise vector, generated cloud scenes have a roughly consistent spatial structure, which suggests that the AI-CD learns from observations that cloud spatial structures are usually determined by other factors than the eight meteorological variables. Meteorological variables nevertheless play a role. For example, cloud spatial structures change substantially due to several variables such as relative humidity, surface pressure, wind-shear, EIS, and subsidence rate. It is also important to note that changes in these variables do not always lead to cloud spatial structure changes, e.g. Figure 3, which suggests that nonlinear combinations between the noise vector and meteorological variables determine what spatial structure of cloud reflectance is like, and that its sensitivity to meteorological conditions are nonlinearly dependent upon noise vectors. Qualitatively, many dependencies discovered by the AI-CD are reasonable based on our understanding of cloud physics. For example, we expect higher cloud fraction and brighter clouds with more of a convective nature as relative humidity increases, which is captured by the bottom row of Figure 4; in a scattered cumulus dominated regime, increase in SST destabilizes the atmosphere and encourages convection, which is captured by the first row in Figure 3. A few extended work will exhaustively and quantitatively verify if such dependencies behave in the same way in real data.

The AI-CD is capable of generating numerous different spatial structures when we change the noise vectors. Most of them have resemblance to corresponding real MODIS scenes although the quality of generated samples is still not as sharp as the real data, which will be discussed further later. This indicates the AI-CD can generate diverse and realistic distributions of 2-D cloud structures, not just simply remembering part of the training data. For example, Figure 3 best resemble scenes from the trade cumulus regime with scattered individual convections and low cloud fraction. In addition, this particular noise vector may contain variables that instruct the AI-CD to generate scenes with sun-glint. Figure 4 best resemble closed-cellular stratocumulus clouds that are starting to transition into either open-cellular or cumulus regimes (H. Wang & Feingold, 2009). Many more spatial structures can be generated when we change the noise vectors.

We have already discussed the sensitivity of cloud spatial structure to meteorological variables and its regime dependency. The sensitivity of cloud fraction generated by AI-CD to meteorological variables are also regime dependent. Cloud fraction is a critical parameter whose sensitivity to meteorological variables is important to model correctly. To illustrate, first we qualitatively examine Figure 3. Cloud fraction strongly increases with SST in this regime which agrees with our expectations. Strong increase in cloud fraction is also associated with increase in EIS, which again is in agreement with our physical understanding since increased EIS means a more stable atmosphere and is more amenable for higher low cloud fraction (Klein & Hartmann, 1993; R Wood & Bretherton, 2006). Interestingly, the AI-CD suggests cloud fraction should decrease with increase in surface wind and wind-shear, which may suggest that increase in wind speed or wind-shear can strengthen mixing of dry air and cloudy air and encourage cloud evaporation. The positive sensitivity of cloud fraction to relative humidity agrees with our expectations. Larger subsidence rate reduces cloud fraction according to the AI-CD in a particular regime ((Figure 3). Such sensitivity of cloud fraction to meteorological variables is quite different from the cloud regime in Figure 4. Here SST increase tends to reduce cloud fraction, which has been modeled in large-eddy simulations and in observational analyses of stratocumulus to trade cumulus transition. Sensitivity of cloud fraction to surface wind speed in Figure 4 is also opposite to that in Figure 3. Here increased wind speed tends to increase cloud fraction. Sensitivity to EIS remains the same with increasing EIS leading to higher cloud fraction. The sensitivity of cloud fraction to relative humidity is also robust. Sensitivity to wind-shear is actually non-linear: first cloud fraction decreases with increasing wind-shear and then increases again although with a different cloud spatial structure.

From qualitative examination of other generated samples, we indeed note that the sensitivity of cloud properties such as cloud fraction and cloud spatial structure to different meteorological variables are generally regime dependent. SST, EIS, and relative humidity are among three variables that have the most robust relationship with different cloud properties. This is worth noting because previous regression analyses of monthly mean and gridded data show similar trends, which may serve as indirect support to what our AI-CD learns directly from observations. It is important to point out that the relationship learned by AI-CD has characteristic time-scale of a day or less while many previous analyses focused on monthly time scales and regional means. As a result, we do expect significantly more diverse relationships between cloud properties and meteorological variables from the AI-CD. It is nevertheless encouraging that AI-CD learns many relationships that are physically sound and regime dependent.

## 4. Discussions and conclusion
### 4.1 Promises, limitations, and future work
We aim to further develop the AI-CD approach as a data driven parameterization because it offers a promising framework to stochastically capture sub-grid cloud spatial structure and bulk cloud properties at the same time. This paper presents the initial results of parameterizing sub-gird cloud reflectance distribution based on a cGAN. They are promising as an initial step because with limited training samples and eight simple grid-mean meteorological variables the AI-CD learns meaningful conditional distributions of cloud spatial structure and sensitivity of bulk cloud properties to different meteorological variables. Furthermore, the GAN based AI-CD is data driven and therefore guarantees a realistic distribution of cloud properties if it is trained sufficiently well, which avoids many shortcomings of current approaches that are based on more idealized models or large-eddy simulations since neither has shown the capability to fully capture cloud distributions. The approach can be easily extended to learn distributions of variables of interest based on detailed physical model simulations that are not directly observable.

One limitation of the approach is that we do not obtain an explicit probability distribution that characterizes the AI-CD. Instead, we only get generated samples from this distribution. Second limitation is that successful training of such models requires careful balance of G and D as well as maintaining the ability of G to generate diverse samples instead of collapsing into a few fixed modes. Third, the generated samples are sometimes blurry or contain other artifacts that make them not realistic enough, which suggests the current AI-CD is still not optimal and has room to improve. In a separate work, we successfully generate much more realistic samples with different model setups (Yuan, 2019), which will help future effort to improve quality of samples generated by the AI-CD.

There are a few straightforward lines of future work to build upon this paper. First, the input meteorological variables can be expanded, both in number of parameters and in the horizontal spatial and the vertical structure. For example, instead of only including data from a grid point, the spatial context of the meteorological conditions can be incorporated. We are working on this and preliminary results are promising. Second, the training sample size can be greatly expanded. This initial effort serves as a proof of concept and expanded training size can make the AI-CD to learn from more data, which usually improves the training. Third, future efforts may be spent on predicting z given an x. That is to learn the latent meaning of individual variables in z, which helps to recover parameters that are important for low clouds. Fourth, the same approach can be extended to include other cloud types. Our initial results are also encouraging.

### 4.2 Conclusion

Our paper adopts the generative adversarial network approach to model distribution of cloud reflectance fields based on MODIS observations. The trained model can generate realistic and diverse cloud scenes. More importantly, it recovers qualitatively reasonable sensitivity of cloud spatial structure and cloud fraction to several meteorological variables. Our approach offers a new framework to create data-driven low cloud parameterizations that can correctly capture low cloud distribution in nature, which may hold the promise to alleviate model biases in low cloud simulation.

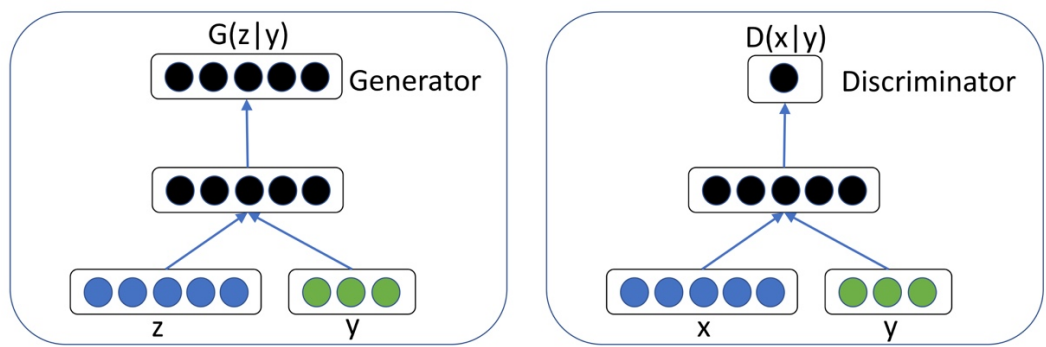

Figure 1: Illustration of the generator and the discriminator in a conditional GAN (a slightly modified version of Figure 1 in Mirza and Osindero, 2014). In this study, y represents the meteorological variables, z the noise vector, and x the image. The generator creates samples from the combination of z and y. The discriminator outputs a true or false vector.

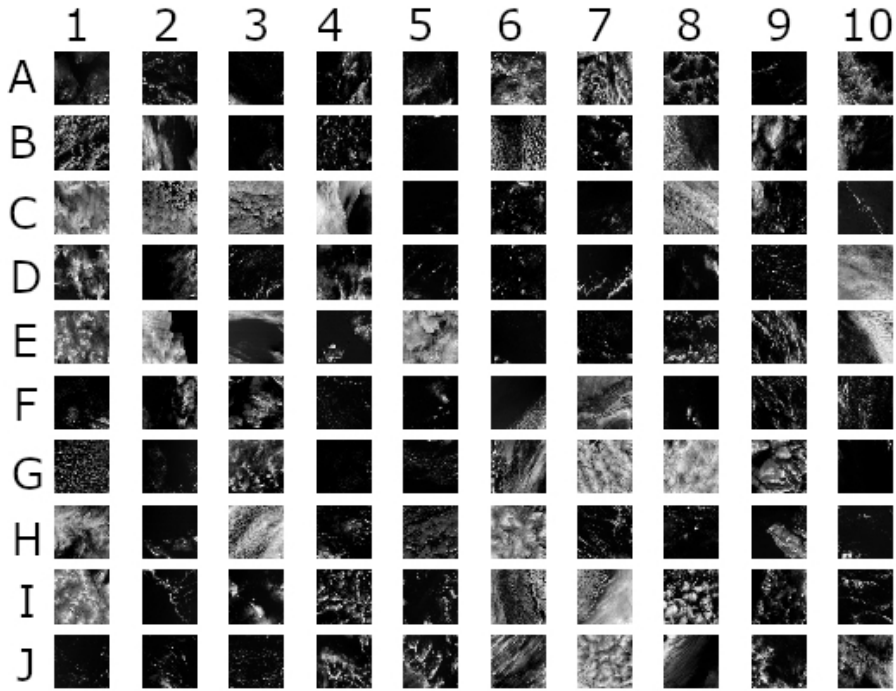

Figure 2: A random selection of 100 real MODIS images of low clouds. Cloud spatial forms display a rich set of possibilities. From scattered convection to open and closed cellular convection. Stratus and other unorganized convection types are also visible.

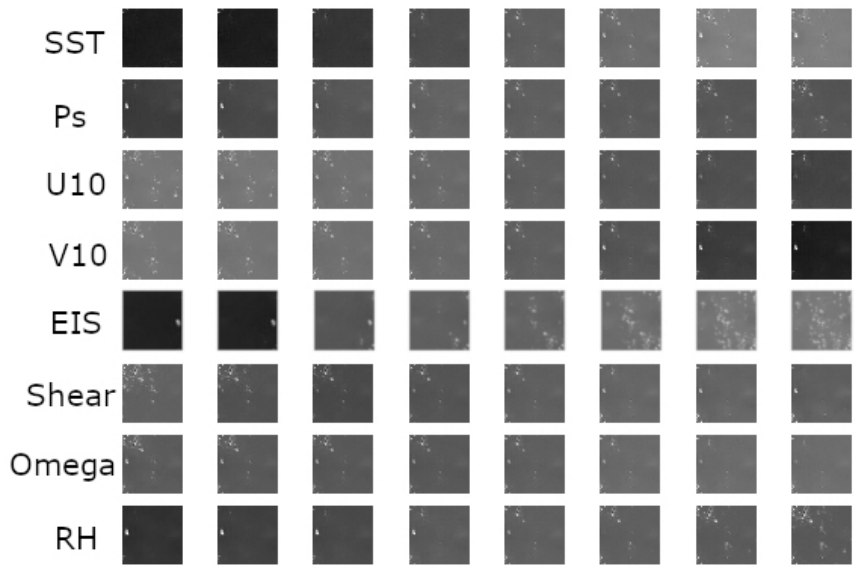

Figure 3. 64 generated samples with the same noise vector. For each row, we only change the row variable while keeping all other meteorological variables constant. Variable values increase from left to right. For this particular figure, the noise vector is configuring a scattered cumulus spatial structure.

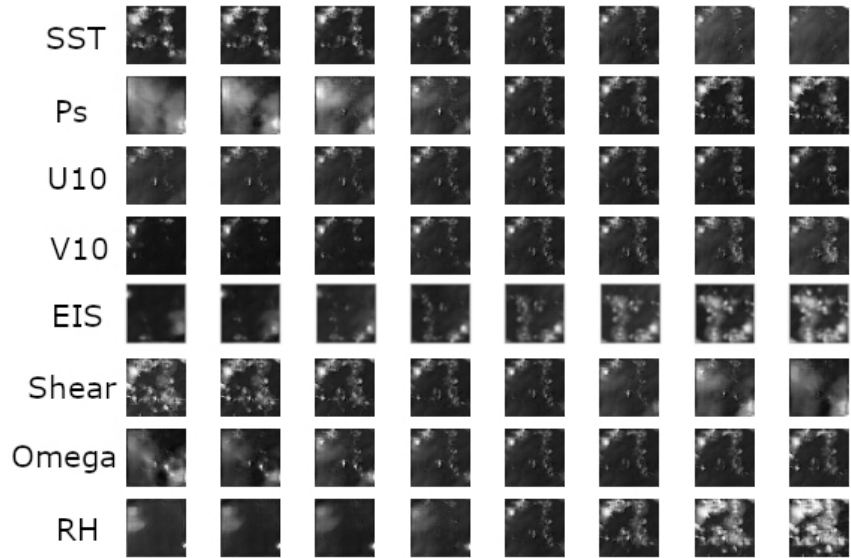

Figure 4. Same as Figure 3 but with a different random noise vector z. The AI-CD is generating an open-cellular cloud form. Changes in meteorological variables can modify it into other forms, indicating both noise vector and meteorological variables having role in deciding the spatial structure. See discussions in the text for more.